\documentclass[12pt,fleqn]{article}

\usepackage{amsmath,amsfonts,bm,amssymb}
\usepackage{amsthm}

\usepackage{ifpdf}

\usepackage{eucal}
\usepackage{eufrak}

\usepackage[margin=2cm,a4paper]{geometry}

\usepackage{url}

\usepackage{multirow}
\usepackage{rotating}

\usepackage{graphicx}

\usepackage{grffile}

\usepackage{color}

\newcommand{\remark}[1]{\marginpar{\scriptsize#1}}
\renewcommand{\remark}[1]{\marginpar{}}   

\long\def\symbolfootnote[#1]#2{\begingroup\def\thefootnote{\fnsymbol{footnote}}\footnote[#1]{#2}\endgroup}

\newcommand{\captionfonts}{\footnotesize}

\makeatletter  
\long\def\@makecaption#1#2{%
  \vskip\abovecaptionskip
  \sbox\@tempboxa{{\captionfonts #1: #2}}%
  \ifdim \wd\@tempboxa >\hsize
    {\captionfonts #1: #2\par}
  \else
    \hbox to\hsize{\hfil\box\@tempboxa\hfil}%
  \fi
  \vskip\belowcaptionskip}
\makeatother   

\newcommand{\be}{\begin{equation}} \newcommand{\ee}{\end{equation}}

\newcommand{\ben}{\begin{enumerate}} \newcommand{\een}{\end{enumerate}}
\newcommand{\bc}{\begin{center}} \newcommand{\ec}{\end{center}}
\newcommand{\bi}{\begin{itemize}} \newcommand{\ei}{\end{itemize}}

\begin{document}
 
\title{Anomalous price impact and the critical nature of liquidity in financial markets}

\author{B. T\'oth\,$^{\textrm{a}}$, Y. Lemp\'eri\`ere\,$^{\textrm{a}}$, C. Deremble\,$^{\textrm{a}}$, J. de Lataillade\,$^{\textrm{a}}$,
\\ J. Kockelkoren\,$^{\textrm{a}}$, J.-P. Bouchaud\,$^{\textrm{a}}$}

\date{\today}
\maketitle
\small
\begin{center}
  $^\textrm{a}$~\emph{Capital Fund Management, 6, blvd Haussmann 75009 Paris, France}\\
\end{center}
\normalsize

\vspace{0.3cm}

\begin{abstract}
We propose a dynamical theory of market liquidity that predicts that the average supply/demand profile is V-shaped and {\it vanishes} around the current price. 
This result is generic, and only relies on mild assumptions about the order flow and on the fact that prices are, to a first approximation, diffusive.
This naturally accounts for two striking stylized facts: first, large metaorders have to be fragmented in order to be digested by the liquidity funnel, 
leading to long-memory in the sign of the order flow. Second, the anomalously small local liquidity induces a breakdown of linear response and a diverging impact of small orders,
explaining the ``square-root'' impact law, for which we provide additional empirical support. Finally, we test our arguments quantitatively using a numerical model of order flow based 
on the same minimal ingredients. 
\end{abstract}

\smallskip




\section{Introduction}
\label{intro}

\begin{figure}
\begin{center}
\includegraphics[width=8cm, height=8cm]{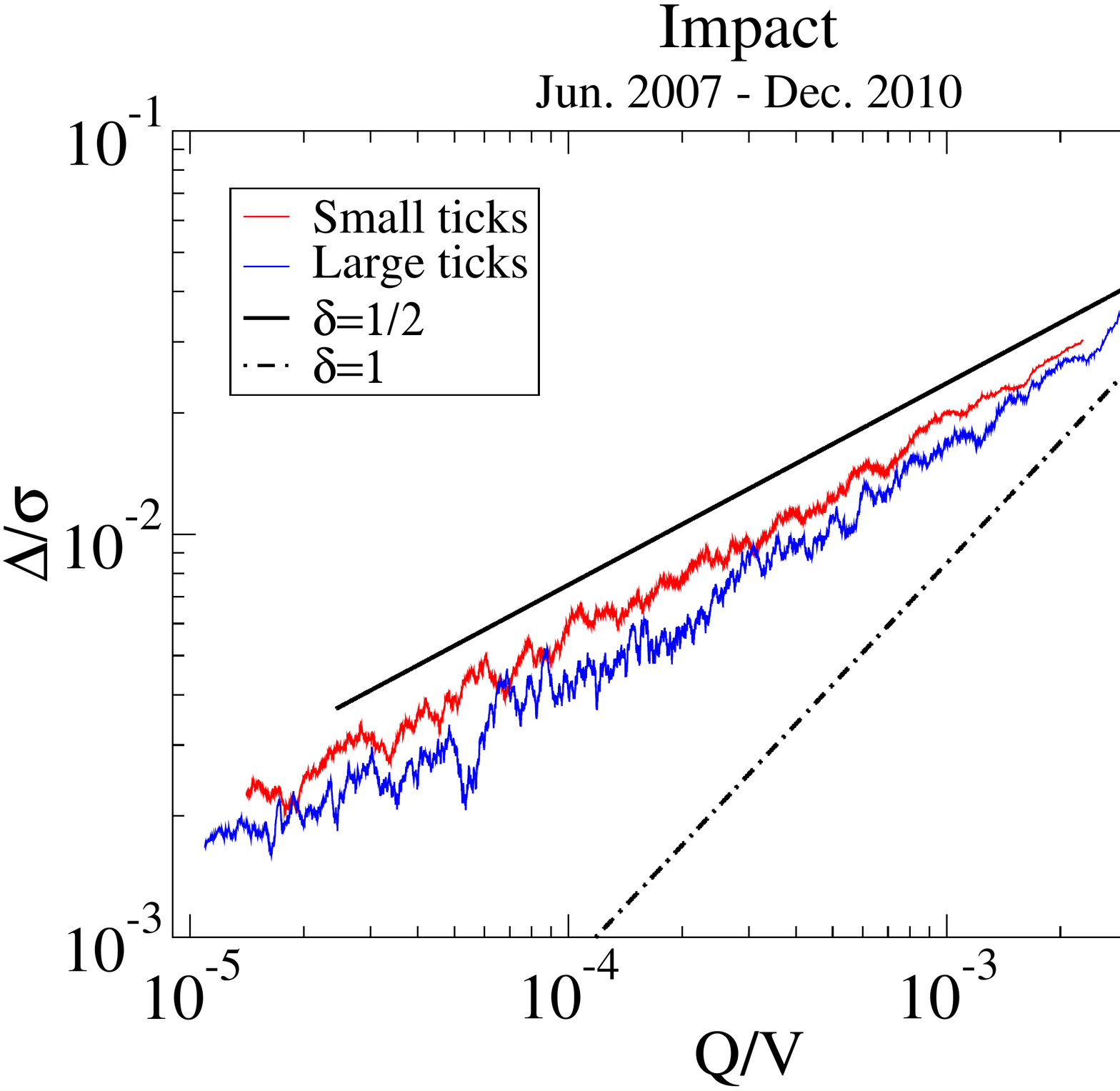}
\caption{\label{fig:data} The impact of metaorders for CFM proprietary trades on futures markets, in the period June 2007 -- December 2010.
Impact is measured here as the average execution shortfall of a metaorder of size $Q$.    
The data base contains nearly 500,000 trades. We show $\Delta/\sigma$ vs. $Q/V$ on a log-log scale, where $\sigma$ and $V$ are the daily 
volatility and daily volume measured the day the metaorder is executed. The blue curve is for large tick sizes, and the red curve is for small tick sizes. 
For large ticks, the curve can be well fit with $\delta=0.6$, while for small ticks we find $\delta=0.5$.
For comparison, we also show the lines of slope $0.5$ (corresponding to a square-root impact) and $1$ (corresponding to linear impact). 
We have removed a small positive intercept $\Delta/\sigma = 0.0015$ for $Q=0$, which is probably due to a conditioning effect.}
\end{center}
\end{figure}

Price impact refers to the correlation between an incoming order (to buy or to sell) and the subsequent price change \cite{Hasbrouck,EQF,BFL}. 
That a buy (sell) trade should push the price up (down) is intuitively obvious and is easily demonstrated empirically (see \cite{BFL} for a recent review). 
Such a mechanism must in fact be present for private information to be incorporated into market prices. 
But it is also a sore reality for large trading firms for which price impact induces extra costs. Indeed, large volumes 
must typically be fragmented and executed incrementally. But since each ``child order'' pushes the price up or down, the total cost of the 
``metaorder''\footnote{{We call the ``metaorder'' (or parent order) the bundle of orders corresponding to a single trading decision. A metaorder is typically
traded incrementally through several child orders.}}
is quickly dominated, as sizes become large, by the average price impact. Monitoring and controlling impact has therefore become one of the most active domain of research in quantitative finance since the mid-nineties. A huge amount of empirical results has accumulated over the years, in particular concerning the relation between the total size $Q$ of the 
metaorder and the resulting average price change. These empirical results come either from proprietary trading strategies (and are often not published), or
from brokerage firms, who execute on behalf of clients \cite{Almgren,Engle,Lehmann,DB,JPM}, or else from the exchanges, who give exceptional access to identification codes that allow one to reconstruct
the metaorders from some market participants \cite{Moro,Menkveld}. Remarkably, although these data sets are extremely heterogeneous in terms of markets (equities, futures, FX, etc.), epochs (from the
mid-nineties, when liquidity was provided by market makers, to present day's electronic markets), market participants and underlying trading strategies (fundamental, technical, etc.) and style of 
execution (using limit or market orders, with high or low participation ratio, etc.), a very similar concave impact law is reported in most studies. More precisely, the average relative price 
change $\Delta$ between the first and the last trade of a metaorder of size $Q$ is well described by the so-called ``square-root'' law:
\be 
\label{sqrt}
\Delta(Q) = Y \sigma \sqrt{\frac{Q}{V}},
\ee
where $\sigma$ is the daily volatility of the asset, and $V$ the daily traded volume, both quantities measured contemporaneously to the trade. The numerical constant $Y$ is of order unity. Published and unpublished data suggest slightly different versions of this law; in particular the $\sqrt{Q}$ dependence is more generally described as a power-law relation 
$\Delta(Q) \propto Q^\delta$, with $\delta$ in the range $0.4$ to $0.7$ \cite{Almgren,Engle,Moro,Lehmann,DB,JPM}. For example, using a large data sample of 700,000 US stock trade orders executed by 
Citigroup Equity Trading, Almgren et al. \cite{Almgren} extract $\delta \approx 0.6$. Moro et al. \cite{Moro} report $\delta \approx 0.5$ for trades on the Madrid stock exchange and $\delta \approx 0.7$ 
for the London stock exchange. We show in Fig. \ref{fig:data} our own proprietary data corresponding to nearly 500,000 trades on a variety of futures contracts, which yields $\delta \approx 0.5$ for small tick 
contracts and $\delta \approx 0.6$ for large tick contracts, for $Q/V$ ranging from a few $10^{-4}$ to a few $\%$. Our data on stocks is also compatible with $\delta \approx 0.5$, although more 
noisy. We note that all these studies differ quite significantly in the details of (a) how the price impact $\Delta$ is defined and measured; (b) how different assets and periods are collated 
together in the analysis; (c) how the fit is performed: over what range of $Q/V$, adding an intercept or not, etc. But in spite of all these differences and those mentioned above -- in particular 
concerning the strategies motivating the trades --, it is quite remarkable that the square-root impact law appears to hold approximately in all cases. 

The aim of the present paper is to provide a theoretical underpinning for such a universal impact law. We first give a general dynamical theory of market liquidity 
that predicts that the average supply/demand profile is V-shaped around the current price. The anomalously small local liquidity induces a breakdown of linear response and
explains the square-root impact law. We then study numerically a stylized model of order flow based on minimal ingredients. The numerical results fully support our analytical arguments, and
allow us to get quantitative insights into various aspects of the problem.

\section{An intriguing impact law}

One should first carefully distinguish the total impact of a given metaorder of size $Q$ from other measures of impact that have been reported in the literature. One is the immediate impact of an 
{\it individual market order} of size $q$, which has been studied by various authors and is also strongly concave as a function of $q$, i.e. $q^\alpha$ with $\alpha \approx 0.2$, or even $\ln q$ \cite{BFL}. 
Another easily accessible measure of impact is to relate the average price change $\Delta_T$ in a given time interval $T$ to the {\it total} market order imbalance ${\cal Q}_T$ in the same time period, 
i.e. the sum of the signed volumes of all market orders. This quantity is estimated using all the trades in the market (i.e. those coming from different market participants) and is clearly different from the 
impact of a given metaorder (see below). However, there seems to be quite a bit of confusion in the literature and many authors unduly identify the two quantities.  
If $T$ is very short, such that there are only one or a few trades, one essentially observes the concave impact of individual orders that we just mentioned.
But as $T$ increases, such as the number of trades becomes large, the relation between $\Delta_T$ and ${\cal Q}_T$ becomes more and more {\it linear} for small imbalances (see e.g. \cite{BFL}, Fig. 2.5), 
and on time scales comparable to those needed to complete a metaorder, the concavity has almost disappeared, except in rare cases when ${\cal Q}_T/V$ is large, in any case much larger than the region where Eq. (\ref{sqrt}) holds. 

A square-root singularity for small traded volumes is highly non-trivial, and certainly not accounted for in Kyle's classical model of impact \cite{Kyle}, which predicts a linear impact $\Delta \propto Q$.  A concave impact function is often thought of as a saturation of impact for large volumes. We believe that the emphasis should rather be placed on the {\it anomalous high} impact of small trades. 
Numerically, Eq. (\ref{sqrt}) means that trading one hundredth of the daily volume moves the price by a tenth of its daily volatility, which is indeed a huge amplification. 
Mathematically, Eq. (\ref{sqrt}) implies that marginal impact diverges for small volumes as $Q^{-1/2}$, meaning that the susceptibility of the market to trades of vanishing size is formally infinite. In most systems, the response to a small perturbation is linear, i.e. small disturbances lead to small effects. The breakdown of linear response often implies that the system is at, or close to, a critical point, 
where very special properties emerge, such as long-range memory or scale invariant avalanches, that accompany this diverging susceptibility. 
Of course, the mathematical divergence is cut-off in practice -- for one thing, the volume $Q$ of a metaorder cannot be smaller than a single lot. Empirical data will never be in the asymptotic limit $Q/V \to 0$, but this is irrelevant to our discussion. This is in fact also the case for most physical systems for which critical behaviour is observed. The important point here is that the proximity of
a critical point can lead to strongly non-linear effects and extreme fragility. As we will argue in detail below, and substantiate within a precise numerical model, we believe that markets operate in a critical regime where liquidity vanishes. This offers a framework to understand many of the anomalies in the behaviour of markets, including the long term memory in order flow and the presence of 
frequent unexplained jumps in prices, that are -- or so we believe -- a consequence of the chronic lack of liquidity that lead to micro-crisis. The anomalous high impact of small trades implied by 
the concave impact law, Eq. (\ref{sqrt}), is, in our view, another side of the same coin. 

Numerous interpretations have been put forth to explain a concave impact law, and can be broadly classified into three types of 
mechanisms (which are not necessarily exclusive): a) risk-reward optimisation of the liquidity providers/market makers, 
\cite{Barra,Grinold,Gabaix-big,Chacko,Zhang}; b) surprise in order flow and decay of impact \cite{BFL,Farmer-new} and 
c) locally linear supply profiles \cite{Harris}. Many of the above stories require that liquidity providers know the fundamental value, carefully monitor the order flow, identify metaorders 
and adjust their quotes such that they eke out some profit or at least break even. There is no doubt that a fraction of market participants strive to achieve such goals, 
and develop astute algorithms in this aim. However, liquidity providing is not (anymore) the monopoly of these market participants, who compete with anyone placing limit orders as part 
of an execution strategy, and the concave impact law in fact holds even when a substantial fraction of the metaorder is executed using limit orders.
That individual metaorders can be detected using statistical methods, as advocated in \cite{Farmer-new}, may well be true for large metaorders of unsophisticated traders, 
but it would be surprising (although not impossible) that our own trades, which are used to obtain Fig. \ref{fig:data}, can be systematically detected. 
The universality of the concave impact law suggests that a robust self-organizing mechanism is at play. Our thesis, that we will substantiate below using both analytical arguments and 
numerical simulations, is that one can indeed build a theory of impact that relies on minimal assumptions, with no reference whatsoever to notions such as fundamental prices, market maker profit, or adverse selection.

\section{A dynamical theory for linear supply/demand profiles}

An interesting idea is that the supply/demand curve is a growing function of the difference between the fundamental value $p_0$ and the price. 
More precisely, suppose the available volume for sells (resp. buys) at price $p$ or above (below), ${\cal V}_\pm(p)$, is a linear function, 
$\pm b_\pm(p-p_0)$. 
The execution of a volume $Q$ of buy orders must therefore take place by moving the price by a quantity $\Delta$ such that:
\be\label{Vshape}
Q = \int_{p_0}^{p_0+\Delta} {\rm d}p \, {\cal V}_+(p) = \frac{b_+}{2} \Delta^2,
\ee
and similarly for sells. Therefore, if the supply/demand profile is {\it linear}, the impact is a square-root: $\Delta_\pm \approx \sqrt{2Q/b_\pm}$. 
It is indeed tempting to wave hands and argue that the available volume is proportional to the mispricing $|p-p_0|$, but one soon hits a major impediment: what exactly is $p_0$? If the above argument is to make any sense, $p_0$ is the fundamental value when the metaorder starts trading, 
and the assumption is that $p_0$ should not vary too much (compared to $\Delta$) during the execution of the metaorder. But this is absurd: there is no equilibrium price around which the market pauses; 
prices move all the time, in a diffusive way, in such a way that difference between $p_0$ and the final price $p_T$ is in fact {\it much larger} than the impact $\Delta$ we try to account for. 
If the linear profile follows instantaneously the price, the above argument completely falls into pieces since only the locally available volume would play a role. 
If the linear profile for some reason remains centred for some time around a specific $p_0$, why should this price coincide with the price at the beginning of the metaorder? 

Still, the basic intuition, that the available volume grows as price excursions get larger, must somehow make sense.  The aim of the present section is to propose a dynamic theory of liquidity largely inspired from \cite{smith03}, based on minimal and plausible assumptions, that indeed predicts that the average supply (or demand) is a V-shaped curve that vanishes around the current price $p_t$. The square-root impact 
then follows from an argument similar to Eq. (\ref{Vshape}). These arguments are then tested quantitatively using a numerical model of order flow based on the same minimal ingredients. 

Our basic idea is that of a ``latent order book'' that at any instant of time $t$ aggregates the total {\it intended} volume for sells at price $p$ or above, ${\cal V}_+(p,t)$,  and the
total {\it intended} volume for buys at price $p$ or below, ${\cal V}_-(p,t)$. We want to emphasize that this is in general {\it not} the volume revealed in the real (observable) order book, in particular 
for $p$ remote from the current price $p_t$. It is rather the volume that {\it would} reveal itself in the order book, or as market orders, if the price came instantaneously closer to $p$. But since there is little incentive to reveal one's intentions too early, 
most of the volume is latent and not revealed. This is obvious from basic order of magnitude estimates: whereas the total instantaneous 
volume in the real order book of a typical liquid stock is of the order of $10^{-5}$ of the market capitalisation, the total transaction volume per day is $10^{-3}$, showing that liquidity is a dynamical process. The empirical analysis of Weber and Rosenow \cite{Weber} shows very clearly how the volume appearing in the order book is indeed stimulated by the trades themselves. 

So our latent volumes ${\cal V}_\pm(p,t)$ reflect intentions that do not necessarily materialize. How do these volumes evolve with time? Between $t$ and $t+ {\rm d}t$, new buyers/sellers may become interested at levels $p_t \mp u$, at a rate $\lambda(u)$ and with unit volume $q=1$; while existing buyers/sellers at $p_t \mp u$ might change their price to $p_t \mp u'$ at rate $\nu(u,u')$, or even disappear temporarily (corresponding to $u' = \infty$).\footnote{The following equations would not change if we allow $q$ to fluctuate, provided the average of $q$ is finite and set to unity without 
loss of generality. We 
furthermore assume complete symmetry between $p > p_t$ and $p < p_t$; i.e. $\lambda_+(u)=\lambda_-(u)$, etc.} Clearly, ${\cal V}_+(p < p_t,t)=0$ and ${\cal V}_-(p > p_t,t)=0$, meaning that there cannot be unsatisfied seller (buyer) below (above) the current price $p_t$.

We now assume that the price process $p_t$ is a Brownian walk\footnote{That the price is a diffusive process is a standard assumption in quantitative finance. It is also very well corroborated by data 
down to very short time scales (see e.g. \cite{subtle}), at least in liquid markets. This ``statistical efficiency'' precludes the existence of simple arbitrage strategies.}, which is only an 
approximation since in practice (a) at short times, microstructure effects play a role and (b) large jumps are present, and in fact quite frequent (the distribution of price changes is well known to be a power law for 
large arguments \cite{plerou}). This approximation however allows us to make precise analytical calculations that illustrate our main point. Since we are interested in phenomena that take place on time scales of a few minutes to a few days, drift effects are completely negligible and we discard them. For the same reason, the difference between an additive and a geometric Brownian motion is irrelevant. A simple equation for the latent volume, averaged over price paths $\overline{{\cal V}_\pm(p,t)}$, can be obtained by working in the reference frame moving with the price $p_t$, provided an extra assumption is made on the rates $\nu(u,u')$. We assume that either $u' = \infty$, with rate $\nu_\infty(u)$, or that the change of price is small, and occurs equally often up or down. We define
$D(u) = \int {\rm d}u' \, (u-u')^2 \nu(u,u')$, where the integral over $u'$ is rapidly convergent (small jumps); it can be interpreted as the (squared) volatility of intentions. With 
${\cal D}(u)=D(u)+\sigma^2$, where $\sigma$ is the price volatility, the final equation for $\rho_\pm(u,t) = \overline{{\cal V}_\pm(p_t \pm u,t)}$ reads \cite{BMP}:
\footnote{Here we assume that $F(u)=\int {\rm d}u' \, (u-u') \nu(u,u') = 0$ $\forall u$, but adding a non zero drift term in the reconfiguration of orders would not change any of the 
following conclusions. Only the value of $u^*$ would change.}
\be
\frac{\partial \rho_\pm(u,t)}{\partial t} = \frac12 \frac{\partial^2}{\partial u^2} \left[{\cal D}(u) \rho_\pm(u,t)\right] - \nu_\infty(u)\rho_\pm(u,t) + \lambda(u); \qquad \rho_\pm(u < 0) \equiv 0. 
\ee
Note that is all rates are symmetric for buy orders and sell orders, the long-time, stationary solution $\rho_{st}(u)$ is the same for $\rho_+$ and $\rho_-$. It describes the most probable shape of the latent order book, and is such that the right hand side of the above equation is zero. For arbitrary function $D(u), \lambda(u)$ and $\nu_\infty(u)$, the explicit form of $\rho_{st}(u)$ is not known, but provided these functions are regular when $u \to 0$, one can show that the stationary profile is {\it linear} close enough to the current price, i.e.  $\rho_{st}(u) \approx b u$ when $u \to 0$, where $b$ is a certain finite constant. In fact, 
\be 
J =  \left. \frac12 \frac{\partial}{\partial u} \left[{\cal D}(u) \rho(u,t)\right] \right|_{u=0}
\ee
is the transaction rate per unit time, i.e. the volume of buy/sell market orders per unit time. If we choose the unit of time to be one day, $J$ is precisely what we called $V$ above. Provided $D(u)$ is regular at $u=0$, the condition $\rho_{st}(u)=0$ and $J$ positive and finite is enough to impose that the profile is locally linear with $b = 2J/{\cal D}(0)$. Therefore, {\it the hypothesis of a diffusive price with a finite transaction rate immediately leads on average to a locally linear order book}.

As a simple illustration, consider the case where new orders appear uniformly, i.e. $\lambda(u) = \lambda$, and ${\cal D}(u)={\cal D}$ independent of $u$. The exact solution is then:
\be\label{solution2}
\rho_{st}(u)= \rho_\infty \left[1 - e^{-u/u^*}\right], 
\ee
with $\rho_\infty = \lambda/\nu_\infty$ and $u^* = \sqrt{{\cal D}/2\nu_\infty}$, leading to $b=\rho_\infty/u^*$. One sees that even when new orders appear with a finite density around the current price, they have also a large probability to be executed and disappear. This eventually leads to a liquidity trough at $u=0$ and a linear profile around $u=0$; $u^*$ can be interpreted as the width of the linear region.
It is reasonable to think that $D$ and $\sigma^2$ are comparable, and that $1/\nu_\infty$, which measures the (volume weighted) average lifetime of an {\it intended} order, is dominated by slow players and is on the scale of a few days.\footnote{Note here that this is where the distinction between latent and revealed orders is crucial: the average lifetime of orders in the order book is much shorter than this, but this is a result of high frequency strategies
which are sensitive to minute price changes, but does not relate to changes of intentions from slow players.  Correspondingly, the average shape of the true order book is non monotonous \cite{BMP}, and thus very different from the linear prediction, Eq. (\ref{solution2}), for the latent liquidity profile.} 
Therefore, $u^*$ is of the order of the daily volatility, which shows that the trough in the latent order book extends over a very significant region 
around the current price. Note that $\tau_\mathrm{life}=1/\nu_\infty$ is also the persistence time of the fluctuations of the order book, and the time to reach the stationary state $\rho_{st}(u)$. This time plays a crucial role in the following.

Note that if $\lambda(u)$ is not constant but decays over a price range $u_\lambda$, the width of the linear region is still given by $u^*$ provided 
$u^* \ll u_\lambda$. In the other limit, on the other hand, one finds $u^* \sim u_\lambda$ \cite{Gatheral}. We believe that $u_\lambda$ and  
$\sqrt{{\cal D}/2\nu_\infty}$ are in fact of the same order of magnitude (a few percent); this means that the players contributing to
the `true' liquidity of the market are not sensitive to price changes much smaller than the daily volatility (see also footnote 3).  

Eq. (\ref{solution2}) contains the central result of the present paper. It predicts that the available volume in the immediate proximity of the current price goes to zero, which is the reason why we say markets are ``critical'', i.e. they operate in a regime of vanishing liquidity. This scenario does not arise by {\it fiat} but is rather a natural consequence of the diffusivity of prices: we believe this is a genuine example Self Organized Criticality \cite{Bak}. 

In more concrete terms, the volume at the best quotes (bid or ask), given by $q_\mathrm{best} \approx b w^2/2$
(where $w$ is the tick size) is $\sim (w/u^*)$ smaller than the volume $\rho_\infty w$ one would expect in the absence of the above sweeping mechanism. This is typically small since $w \sim 0.05 \%$ and $u^* \sim 2\%$ (for stocks or futures).  One can also compare $q_\mathrm{best}$ to the the total volume traded in a time $T$, which is $V =JT \sim \rho_\infty u^*$ if one chooses $T = \tau_\mathrm{life} \sim$ one 
day. The ratio is now $q_\mathrm{best}/V\sim (w/u^*)^2$, which is very small, as is indeed the case empirically: the immediately accessible volume is typically a 1000 smaller than the daily turnover. This small liquidity 
compels market participants to fragment their trades, leading to the universally observed long-range correlation in the sign of market orders \cite{subtle,LF}. It also leads to the square-root impact law, if one trusts the argument after Eq. (\ref{Vshape}), with here $b=\rho_\infty/u^*$. This gives $\Delta \approx \sqrt{\frac{2Q}{b}} \propto \sigma \sqrt{\frac{Q}{J}}$, where we have used $b \sim J/\sigma^2$, with $J$ is the trading rate. This is exactly the square-root impact law, Eq. (\ref{sqrt}). 

However, this last argument is quite na\"ive since the average impact $\Delta$ is much smaller (for small $Q$) than the typical excursion of the price within the same period. Furthermore, the diffusive behaviour of the price is taken for granted in the above calculation, whereas in fact it results from a subtle compensation \cite{subtle} between a {\it confinement effect} created by the linear supply/demand curve (a price movement in one direction hits larger opposing volumes and is more likely to revert) and a {\it correlation effect}, created by the fragmentation of the trades \cite{LMF,toth2011} -- itself imposed by the liquidity trough mentioned above. 

Since we are unable at this stage to treat these effects consistently within an analytical framework, we now turn to a minimal numerical model that captures all the above effects. We will indeed find a linear demand profile and a concave impact function, and gain considerable insights into the mechanisms leading to such behaviour. 

\section{A numerical model for an efficient market with long-ranged order flow} 

The numerical implementation of the above simple Poisson model for intended order flow is quite simple, and we follow the framework proposed in \cite{smith03} -- see also the Appendix below. 
All orders have unit volume. Limit orders are launched at a uniform rate $\lambda$ in a finite 
(but large) interval around the current price. Existing limit orders are individually cancelled at rate $\nu_\infty$. If market orders are themselves completely random, with buys and sells drawn independently with probability 
$\frac12$ and at rate $\mu$, the result price motion is known to be strongly {\it subdiffusive}, in the sense that the lag dependent diffusion constant 
$\sigma^2(\ell) \equiv \langle (p_t - p_{t+\ell})^2 \rangle/\ell$ decays when $\ell$ increases (here time is counted in number of transactions). This was noted 
in \cite{smith03}, and is a result of the confining effect of the supply/demand curve. The price only becomes diffusive on time scales larger than the memory time 
$\tau_\mathrm{life}$. Nothing of that sort is seen in real price dynamics. However, we know from empirical data \cite{subtle,LF,BFL} that the signs of transactions $\epsilon_t$ are in fact long-ranged correlated, 
i.e. that $C(\ell)=\langle \epsilon_t \epsilon_{t+\ell} \rangle$ is decaying as a power-law, $C(\ell) \approx \ell^{-\gamma}$, where $\gamma \approx 0.5$ for single 
stocks and $\gamma \approx 0.8$ for futures markets \cite{lemperiere_unpublished}. This persistent direction of trading can counterbalance the confinement effect and restore diffusion. In fact, 
if $\gamma$ is too small, one expects {\it superdiffusion}, i.e. $\sigma^2(\ell)$ growing with $\ell$. 

We therefore want to upgrade the ``zero-intelligence'' model of \cite{smith03} to an ``$\varepsilon$-intelligence'' numerical model which explicitly includes
the long ranged correlated nature of the trades, 
reflecting the fact that large metaorders are fragmented and traded incrementally, see Ref. \cite{LMF,toth2011}. We have chosen to work with the LMF specification of the sign 
process \cite{LMF}, i.e. sequences of $+/-$ signs (buy/sell market orders) are generated, with lengths $L$ drawn from a power-law distribution: $P(L) \sim L^{-(\alpha+1)}$. 
It can be shown that the sign process has an 
autocorrelation function that decays asymptotically as $C(\ell)=\ell^{-\gamma}$ with $\gamma=\alpha-1$. We do not expect that the following results depend much on the precise 
specification of the model of signs \cite{usinprep}.

At this stage, the model contains four parameters: the rate of limit orders $\lambda$, the rate of cancellation $\nu_\infty$, the rate of
market orders $\mu$ and the autocorrelation exponent $\gamma$. In fact, only three of them are relevant, up to a change of time unit: the ratios $\lambda/\nu_\infty$ and $\mu/\nu_\infty$, and $\gamma$. 
As noted above, the ratio $\rho_\infty=\lambda/\nu_\infty$ determines the supply depth far
away from the best quotes, whereas the ratio $r=\mu/\nu_\infty$ tells us whether we are in a ``slow'' market, where the renewal time $\tau_\mathrm{life}$
of the supply/demand is much longer than the time between individual trades ($r \gg 1$), or if the market is ``fast'' in the opposite limit $r \ll 1$. It is clear that real markets are in the former limit: if trades take place on a scale of seconds while the renewal time is of the order of hours or days, the ratio $r$ is on the order of $10^4$. As will be discussed later, $r \gg 1$ is a crucial condition for a concave impact law to hold. On the other hand, $\mu$ and $\lambda \times w$ are of similar order of magnitude, which means that, as expected, markets are also ``deep'': $\rho_\infty \times w \gg 1$. 

The problem is that in the limit of deep and slow markets, the above model is always in the subdiffusive phase unless $\gamma$ is very close to zero (see Fig. \ref{fig:diffusivity} below). We therefore need an extra ingredient to make markets statistically efficient (meaning that prices are diffusive), while keeping $\gamma$ in the empirical range $\gamma \sim 0.5 - 0.8$. When compared with real markets, the above model is obviously much too simple. For example, the size of both limit and market orders is known to be broadly distributed, whereas we assume, as in \cite{smith03,cont11}, that all volumes are unity. The direction and the size of market orders are furthermore strongly conditioned by the volumes at the best quotes $q_\mathrm{best}$: the volume 
of market orders is larger when the offered volume is 
larger, and the sign of the next market order is anti-correlated with the volume imbalance (i.e. when $q_\mathrm{bid} \gg q_\mathrm{ask}$, the next
trade is most likely to be a buy, and vice-versa). 

We have chosen to add one extra parameter both to make a step towards reality and solve the 
efficiency problem, in the following way: the volume of market orders is chosen to be a random fraction $f$ of the opposite volume at the best, where
the distribution of $f$ is given by $P_\zeta(f)=\zeta (1-f)^{\zeta-1}$, where $\zeta$ is a parameter ($\zeta>0$), that determines the typical relative volume of market orders. For $\zeta\rightarrow 0$, the 
distribution $P_\zeta(f)$ peaks around $f=1$ and most orders ``eat'' all the available liquidity; $\zeta=1$ corresponds to a flat distribution for the fraction
of eaten volume; finally the limit $\zeta\rightarrow\infty$ corresponds to very small (unit) volumes and recovers the previous model. The correlation between
volume at best and volume of the impinging market order has been reported in many papers (see \cite{farmer2004}). This makes perfect sense: since 
large metaorders must be fragmented because of the small available liquidity, one expects that the executed volume opportunistically follows the 
offered liquidity. 

With the help of the extra parameter $\zeta$ we can now tune the model to guarantee diffusive prices for any value of $\gamma$, even in the 
limit of deep and slow markets. We define a measure of pure diffusivity as $\sigma(\ell_1)/\sigma(\ell_2)$ for two time scales $\ell_1$ and $\ell_2
> \ell_1$. Subdiffusion corresponds to ratios $< 1$, superdiffusion to ratios $> 1$ and for pure diffusion this ratio must be equal to unity for 
all $\ell_1, \ell_2$. In our simulations, we chose $\lambda=0.5$, $\mu=0.1$, $\nu_{\infty}=10^{-4}$ (corresponding to $\rho_\infty = 5000 \gg 1$ and 
$r = 1000 \gg 1$. We determine the
{\it diffusion line} in the plane $\gamma, \zeta$, 
that separates the subdiffusion regime from the superdiffusion regime, by setting  $\ell_1=10$, $\ell_2=1000$.\footnote{The 
results are not sensitive to the precise choice of  $\ell_1$ and $\ell_2$, provided $\ell_1,\ell_2 \ll \tau_\mathrm{life}$. We have explored a wide range of values of 
$\lambda, \mu, \nu_\infty$. Provided one remains in the limit of deep and slow markets, the broad picture is unaffected, although the precise location of the diffusion line in Fig. \ref{fig:diffusivity} is changed.} The result is shown in Fig. \ref{fig:diffusivity}. As expected, smaller $\gamma$'s favor superdiffusion, whereas larger values of $\zeta$ (corresponding to less aggressive market orders) 
favors subdiffusion. On the boundary between the two regimes we find the purely diffusive, efficient markets we are looking for; in the following 
we fix the value of $\zeta(\gamma)$ such as to be exactly diffusive.\footnote{It would be interesting to obtain an analytical form for the diffusion 
line, but we have not attempted to make a theory for this at the present stage.} Since the value of $\zeta$ is fixed by the condition of price diffusivity, our model still has three parameters: 
depth ($\rho_\infty$), inverse speed ($r$) and trade persistence ($\gamma$).

\begin{figure}
\begin{center}
\includegraphics[width=8cm, height=5.5cm]{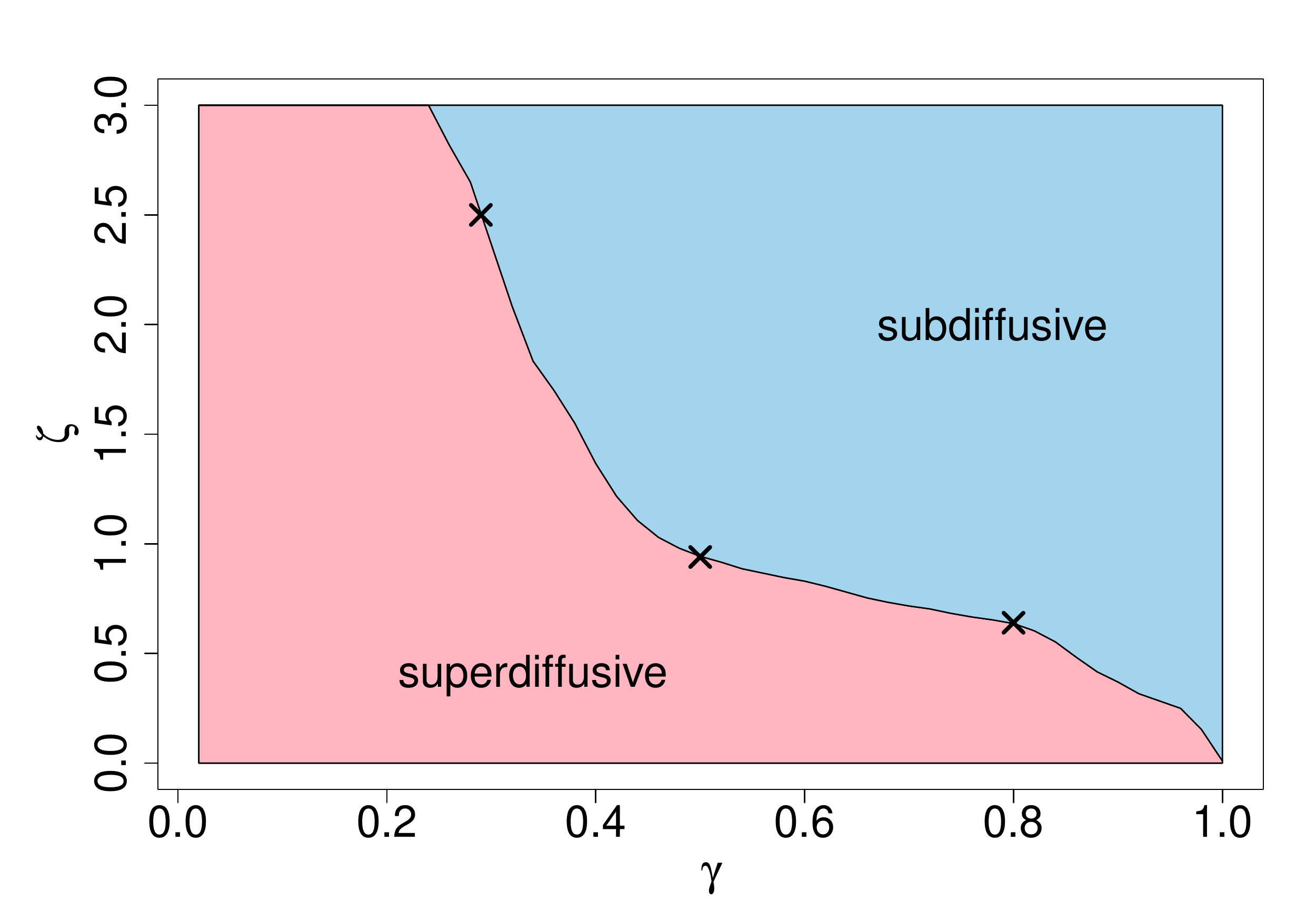}
\caption{\label{fig:diffusivity} Map of diffusivity in the plane $\gamma, \zeta$ for $\lambda=0.5$, $\mu=0.1$, $\nu_{\infty}=10^{-4}$.
The large $\gamma$, large $\zeta$ region corresponds to subdiffusion, while the small $\gamma$, small $\zeta$ region corresponds to superdiffusion.
The ``efficient market'' (diffusive) line determined numerically is such that $\zeta(0.8) \approx 0.65$, $\zeta(0.5)  \approx 0.95$, and $\zeta(0.3)  \approx 2.5$ (crosses). 
Note that the unit volume limit $\zeta \to \infty$ corresponds to $\gamma \to 0$. Without volume fluctuations, deep and slow markets are therefore always 
found to be in the subdiffusive phase.
}
\end{center}
\end{figure}

\begin{figure}
\begin{center}
\includegraphics[width=8cm, height=5.5cm]{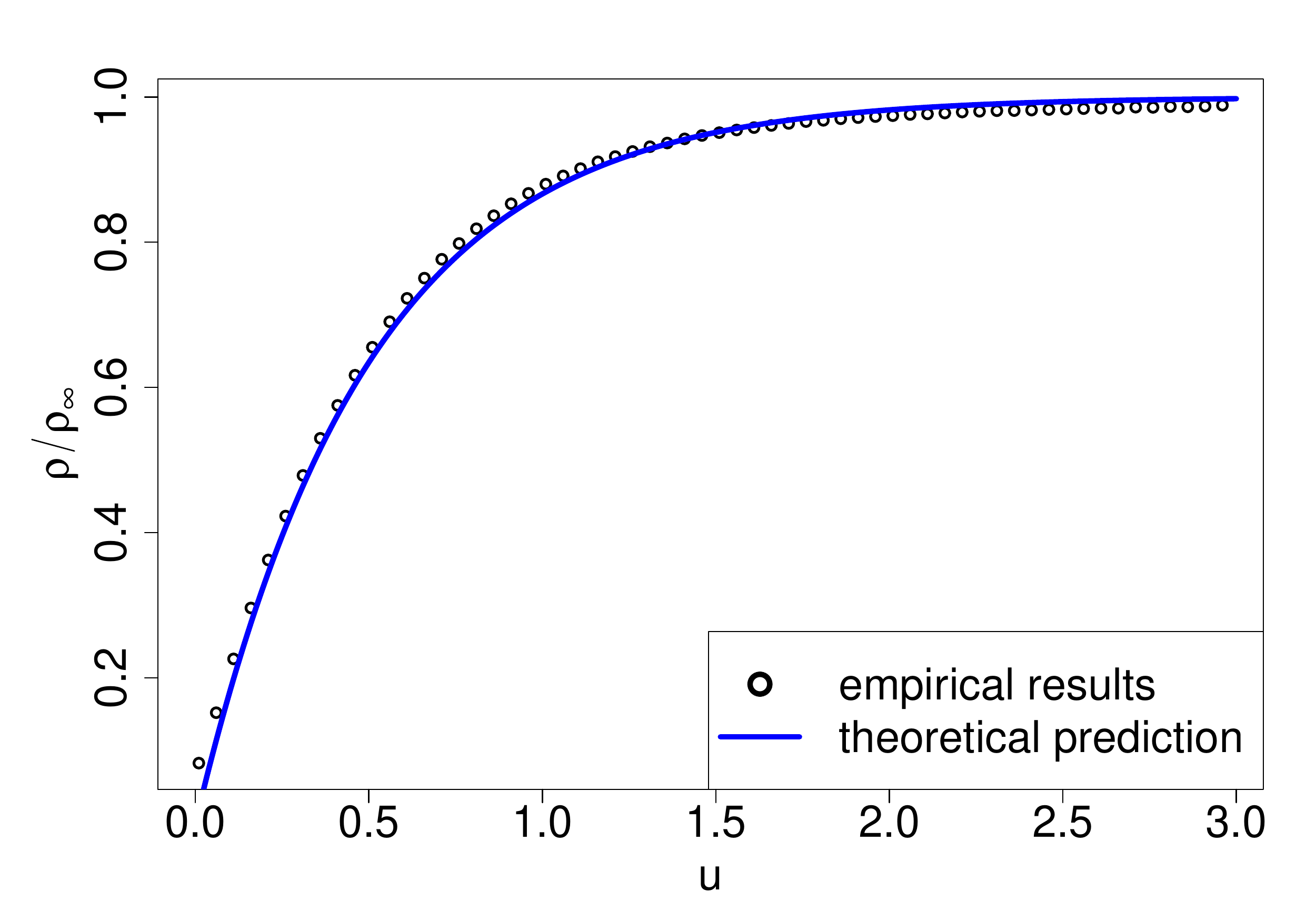}
\caption{\label{fig:extraplot} Comparison of the stationary density of the book with the expected shape calculated from Eq. (\ref{solution2}). The symbols show the empirical shape of the book from the simulations, while the solid line shows the predicted exponential form, with $u^*$ being computed from the actual price volatility.
In the simulations we used the parameters, $\gamma=0.8$, $\zeta=0.65$, $\nu_\infty=10^{-4}$, $\lambda=0.5$. The analytical computations lead to $u^*=\sqrt{D/2\nu_{\infty}}=0.49$, while a direct fit to Eq. (\ref{solution2}) leads to $u^* \approx 0.48$.}
\end{center}
\end{figure}

To illustrate the tight connection between the dynamical theory discussed in the previous section and the above described numerical model, in Fig. \ref{fig:extraplot} we compare the stationary density of the book to its expected shape calculated from Eq. (\ref{solution2}). The symbols show the stationary density, $\rho_{st}$, measured in the simulations, while the solid line is Eq. (\ref{solution2}) with $u^*=\sqrt{D/2\nu_{\infty}}$, $D$ being the actual measured price volatility.
The latent order book is  found to be linear in the immediate vicinity of the price, as predicted. The exponential curve with the analytical expression for $u^*$ leads to a very good fit in the whole range of $u$.

\begin{figure}
\begin{center}
\includegraphics[width=8cm, height=5.5cm]{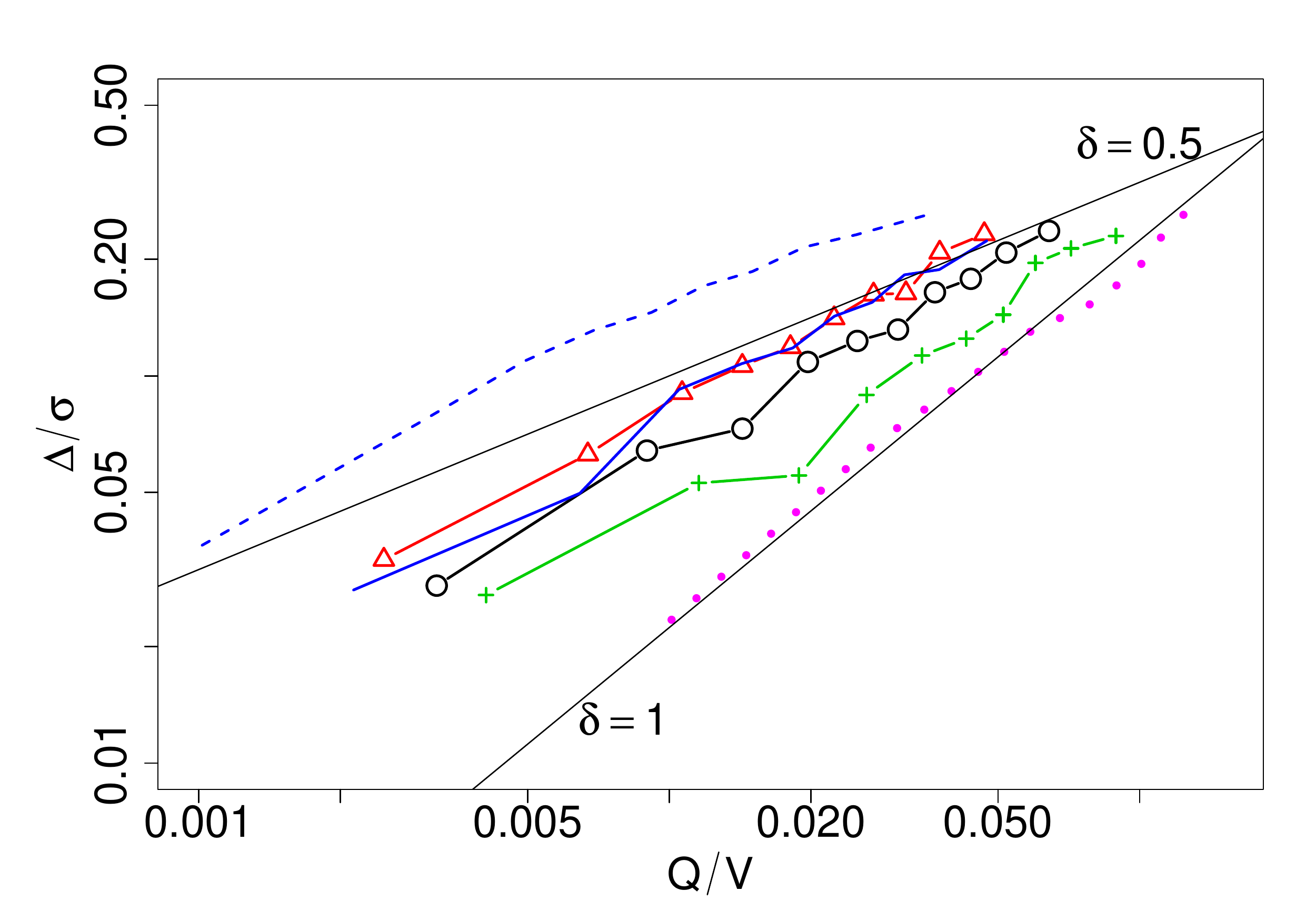}
\caption{\label{fig:slippage} Impact of metaorders on log-log scale. We show $\Delta/\sigma$ vs. $Q/V$, where $\sigma$ and $V$ are 
measured over the timescale $\tau_\mathrm{life}$. The three curves with symbols show the impact for the $\zeta$-execution with high participation rate, 
$\Phi=0.3$, and three values of the sign autocorrelation exponent: $\gamma=0.3$ (green crosses), $\gamma=0.5$ (black circles),
$\gamma=0.8$ (red triangles). The two blue lines show the impact for low participation rate $\Phi=0.05$, for the $\zeta$-execution (full line)
and unit-execution (dashed line), both for $\gamma=0.5$. The purple dots show a case when the time to complete metaorders is longer than the 
lifetime of the book $T>\tau_\mathrm{life}$. In this case we get back a linear impact. The two solid black lines are of slope $\delta=0.5$ and $\delta=1$, for comparison, with the choice $Y=1$ for $\delta=0.5$.}
\end{center}
\end{figure}

We now have a model such that (i) the order flow has long range memory but (ii) the price is diffusive. We are thus in a position to test quantitatively
our above ideas about the linear supply function and the impact of metaorders. In order to do this, we add to the above order flow an extra agent who becomes active at a certain (arbitrary) time $t$, chooses a random sign $\epsilon$ and a random size $Q$ for his metaorder, which he executes incrementally
using market orders until all the volume is transacted. We have considered two execution styles: a) ``$\zeta$-execution'', where the extra agent trades 
exactly as the rest of the market, by sending a market order of volume $f \times q_\mathrm{best}$, where $f$ is chosen according to $P_\zeta(f)$ above;
b) ``unit-execution'', where the market orders are all of unit volume, whenever he trades. (We also studied limit order execution with similar results, see \cite{usinprep}.) 
In both cases, he participates to a fraction $\Phi$ of all market orders. We measure, as the real data shown in Fig. \ref{fig:data}, the impact $\Delta$ as the 
price paid by the extra agent compared to the price $p_t$ at the beginning of the metaorder, averaged over many different metaorders of size $Q$. The results are shown in Fig. \ref{fig:slippage}, where we show in a log-log scale $\Delta/\sigma$ on the y-axis versus $Q/V$ on the x-axis, for different values of $\gamma$, $r$ and $\Phi$, and for both $\zeta$-execution and unit-execution. We also show two straight lines of slope $\delta=1/2$, corresponding to a square-root impact, and $\delta=1$, corresponding to a linear impact. It is clear that for all cases where the execution time $T$ is much smaller than the renewal time $\tau_\mathrm{life}$, the impact is strongly concave and, {\it in a first approximation}, independent of $\gamma$ and of the participation rate $\Phi$, in agreement with empirical observations. The exponent $\delta$ is found to be close to $1/2$ for unit-execution, and close to $2/3$ for
$\zeta$-execution. The $Y$ factor, defined in Eq. (\ref{sqrt}), can be measured from the data shown in Fig. \ref{fig:slippage}, and is of order unity, as found empirically. More details about these results will be given in \cite{usinprep}.

\begin{figure}
\begin{center}
\includegraphics[width=8cm, height=5.5cm]{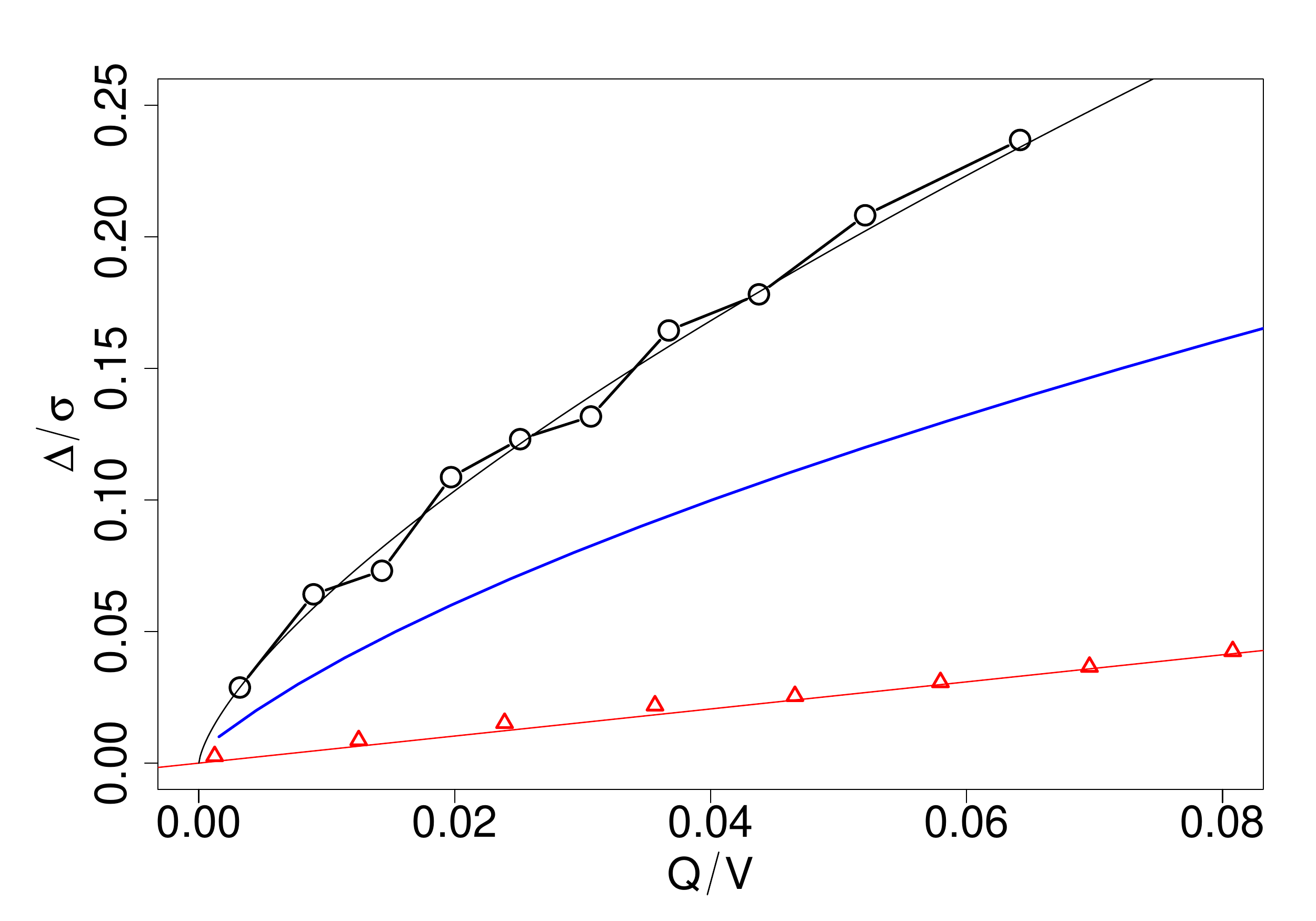}
\includegraphics[width=8cm, height=5.5cm]{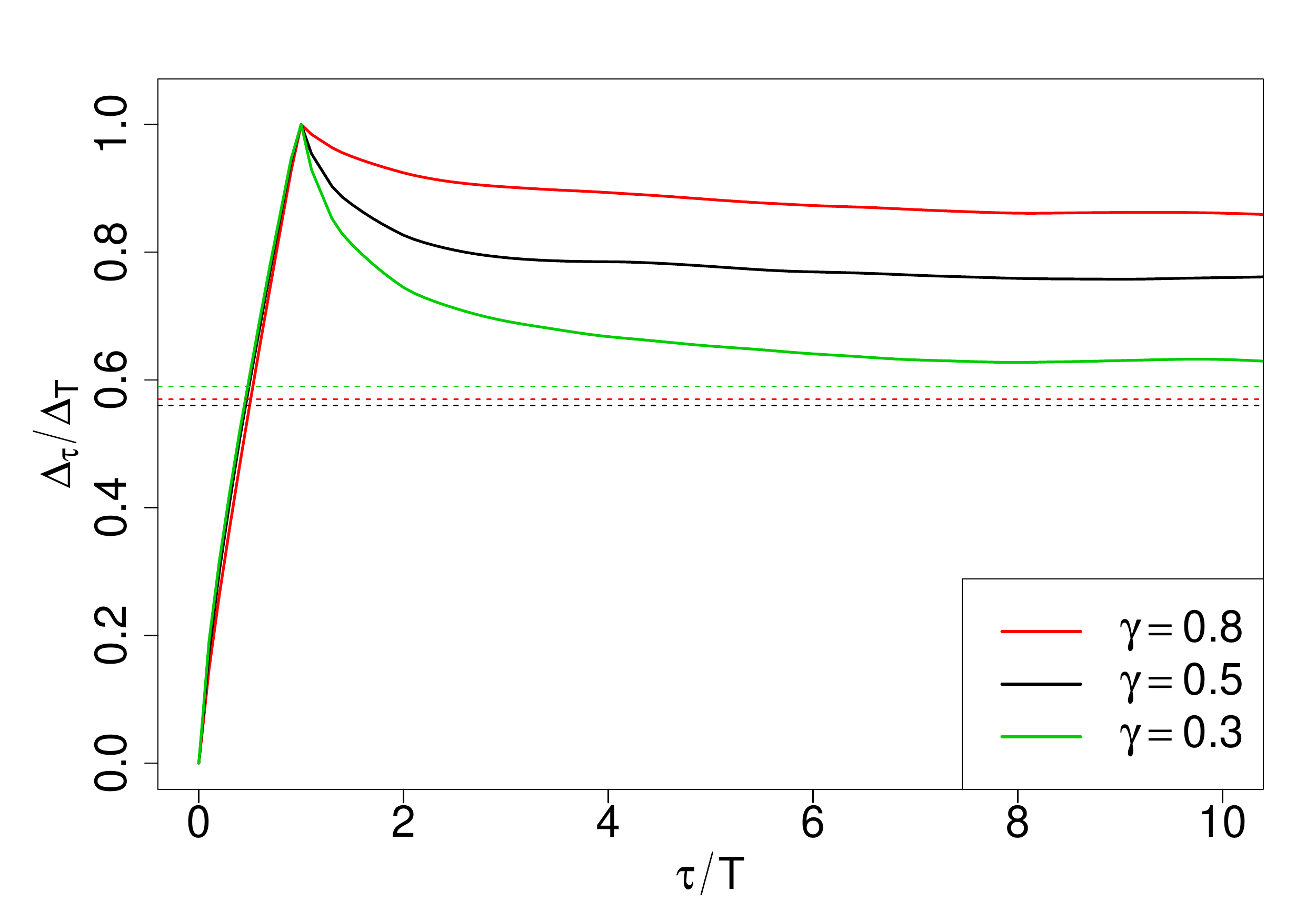}
\caption{\label{fig:4} {\it (Left)} The true impact of $\zeta$-executed metaorders for $\gamma = 0.5$, $\Phi = 0.3$ (black circles), the na\"ive estimate of the 
impact from integrating the volume in the average latent order book (blue solid line), 
and the average price change against the global volume imbalance in a given period (red triangles).
We also show a power law fit for the metaorder impact, as $Y (Q/V)^\delta$, with $\delta=0.7$ and $Y=1.59$ (solid black line) and a linear fit for the global measure of impact (solid red line).
{\it (Right)} The decay of impact after the completion of the metaorder for unit-execution for $\gamma=0.3, 0.5$ and $0.8$ and $\Phi=0.5$. The x-axis is $\tau/T$, the clock-time in units of 
the time needed to complete the metaorder (with $T \ll \tau_\mathrm{life}$), the y-axis is the rescaled price change during the metaorder. After the metaorder, the price appears to relax to a 
constant level, which is a $\gamma$ dependent fraction of the temporary impact. The dotted horizontal lines correspond to the actual average execution price of the metaorder, which is found
to be $\approx 0.6 \, \Delta_T$ for all three values of $\gamma$.}
\end{center}
\end{figure}

Some parameters of the model however do influence the value of the effective exponent $\delta$, which might explain why empirical data show some scatter around the value $\delta=\frac12$. In particular, when the execution time $T$ increases and becomes comparable to the renewal time $\tau_\mathrm{life}$, the effective exponent $\delta$ increases and the impact becomes linear in the 
limit $T \gg \tau_\mathrm{life}$ -- see Fig. \ref{fig:slippage}. This is indeed expected since impact is necessarily additive when all memory of the latent order book has been erased. It is also a direct proof that the persistence of the supply/demand is central to explain the functional form of the impact. 
The na\"ive prediction for the price impact, Eq. (\ref{Vshape}), based on the average slope of the supply curve, is shown in Fig. \ref{fig:4} (left), together with the numerical determination of $\Delta(Q)$ already plotted in Fig. \ref{fig:slippage}. We see that the na\"ive argument 
indeed leads to the correct order of magnitude for the impact, but fails to be quantitatively accurate: it underestimates the real impact by a factor $\sim 2$. On the same graph, we also show the global measure of impact mentioned above, where the average price change is plotted against the total volume imbalance ${\cal Q}$. We see that the latter is very different from the impact of a given metaorder. 
The global impact is {\it linear} in ${\cal Q}$ (as observed on empirical data for large enough $T$) and, for small volume imbalances, much smaller than the impact of an additional metaorder that perturbs the equilibrium flow. 

Another interesting aspect of the problem is the temporal structure of the impact of metaorders. In Fig. \ref{fig:4} (right) we show the average difference 
$\Delta_\tau$ between the price $p_t$ of the first trade of a metaorder and the price $p_{t+\tau}$ at time $\tau$ later, in the direction of the trade. The x-axis is in units of the time needed to complete the metaorder. The exact shape of the build-up in time depends on the execution mechanism, this we will discuss in details in \cite{usinprep}. Here we focus on the price dynamics after the completion of the metaorder.
Once the metaorder is completed (i.e. for $\tau > T$), the impact decays down to smaller values, and appears to reach a plateau $\Delta_\infty$, such that $\Delta_\infty/\Delta_T$ depends on the value of $\gamma$, and is $\sim 0.75$ for $\gamma=0.5$. The cause of the reversion in our model is that during the trading of a metaorder, the order book becomes on average unbalanced: orders on the opposite 
side become statistically more numerous, as a consequence of the V-shaped liquidity. This results in a partial reversion of the price, once the pressure from the metaorder is over.

Interestingly, a similar 
behaviour has also been observed in \cite{Moro}, and is predicted by theory of impact recently put forth in \cite{Farmer-new}, 
which elaborates on the idea of a ``fair execution price''. As emphasized in \cite{Moro,Farmer-new}, if a metaorder of size $Q$ 
has an impact that grows as $(q_{\mathrm{exec}}/Q)^\delta$ (where $q_{\mathrm{exec}}$ is the volume already executed, i.e. $q_{\mathrm{exec}}/Q$ is ``volume time'')\footnote{Since the impact is $\Delta(Q) = Y \sigma (Q/V)^\delta$, we expect the impact to grow as $\Delta_{q_{\mathrm{exec}}} \approx  \Delta(Q) (q_{\mathrm{exec}}/Q)^\delta$ meaning that a metaorder of size $Q$ stopped half-way through execution impacts the price the same way as a metaorder of size $Q/2$, 
which {\it a priori} makes sense since there is no way to anticipate when a metaorder is going to stop. Indeed, we find the build-up of the 
impact in {\it volume} time to be concave.}, then the average execution price of the metaorder is $\Delta_T/(1+\delta)$. In this case, $\Delta_\infty/\Delta_T \equiv 1/(1+\delta)$ ensures that the average execution price of the metaorder is equal to the price long after the execution is completed, in other words that the execution price is fair. Used together with the condition that the price should be a martingale and
that the size distribution of metaorders is a power-law, $Q^{-1-\alpha}$, Farmer et al. \cite{Farmer-new} obtain a concave impact $\Delta \sim Q^\delta$ with 
$\delta=\alpha-1=\gamma$.\footnote{Note that in the theory of Farmer et al. \cite{Farmer-new}, the exponents $\delta$ and $\gamma$ are equal, whereas in our model $\delta$ is to a large 
extent unaffected by the precise value of $\gamma$.} Using $\alpha \approx 3/2$ from empirical data on stocks, the square-root impact law is recovered \cite{Farmer-new}, with  
$\Delta_\infty/\Delta_T = 2/3$. We see however in Fig. \ref{fig:4} (right) that the ``fair price'' condition does not hold in general within our model, since the plateau value 
significantly changes with $\gamma$, whereas the average price paid is close to $0.6 \, \Delta_T$ in all cases. Nevertheless, for $\gamma=0.5$ the plateau is not far from the value $2/3$ predicted by
Farmer et al. \cite{Farmer-new}.

\section{Conclusion}

Although some elements of our model are common with the framework of \cite{Farmer-new} (broad distribution of metaorders leading to long range 
correlation of the order flow, and the efficient price condition), others are very different. The theory advocated in \cite{Farmer-new} requires that 
metaorders can be identified by market makers, in particular that the very first trades can be detected. As the authors openly admit, the need for market 
participants to be able to detect the starting and stopping of a given metaorder is potentially a problem, and in fact our model does not rely at all on 
such a strong assumption. On the contrary, we have shown that the universally observed concave
impact law is a consequence of some robust, generic assumptions about market dynamics. 
In particular, we have provided a dynamical theory 
of liquidity which leads to a locally linear supply/demand curve, provided high frequency strategies guarantee price diffusivity on all time scales. 
Our story is purely statistical and does not rely on additional (and sometimes wooly) notions such as fundamental prices, adverse selection or fair pricing. 
There are no explicit market makers, strategic players or optimising agents in our picture, but rather an ecological equilibrium of indistinguishable traders 
that interact in a way to make the price statistically efficient. 

Although our ``$\varepsilon$-intelligence'' numerical model makes an important step towards reality (in particular by including long-range correlations in the order flow and
ensuring a diffusive price dynamics), there are still many assumptions that are ad-hoc. One knows for example that the deposition/cancellation rates of limit orders do 
strongly depend on the distance from the current price \cite{more}, that the assumption of a Poisson process is an oversimplification \cite{JEDC} (these observations apply to the real order book, but probably will also hold for the latent order book), that the volume of limit orders is not at 
all constant, etc. etc. Building a detailed, realistic model of order flow is of course needed to get fully quantitative predictions for the impact, but was not the scope of the present work. 
We instead wanted to have a simplified model that would allow us to test our central hypothesis: that the latent order book is locally linear and that this is the crucial ingredient to explain
the square-root impact law. We believe that this objective has been reached, and we leave the more ambitious project of building a full-scale model for future work. We also note 
that our central assumption of a latent order book has other, empirically testable assumptions, which we are currently investigating.

As we emphasized in the introduction, understanding impact has immediate practical implications in terms of trading costs and capacity 
of strategies. In our view, the most important message of the theory presented above concerns the critical, inherently fragile nature of 
liquidity. By necessity, a diffusive price leads to a vanishing liquidity in the vicinity of the current price. This naturally accounts for
two striking stylized facts: first, large metaorders have to be fragmented in order to be digested by the liquidity funnel, 
which leads to long-memory in the sign of the order flow. Second, the anomalously small local liquidity induces a breakdown of linear response and a diverging impact of small orders. 
Furthermore, liquidity fluctuations are bound to play a crucial role when the average liquidity is small, and we expect these fluctuations to be at the 
heart of the turbulent dynamics of financial markets, as postulated in, e.g. \cite{Farmer-volume,subtle,Bouchaud2010}.

\section*{Acknowledgements} We want to thank Paul Besson, R\'emy Chicheportiche, Zolt\'an Eisler, P\'eter Horvai, Fabrizio Lillo, 
Matteo Marsili, Marc Potters, Emmanuel S\'eri\'e, Damien Trouv\'e and especially Doyne Farmer and Jim Gatheral for very helpful discussions and suggestions.


\appendix

\section{Appendix}

This Appendix summarizes the main steps of the numerical model discussed in the paper.
In the model we consider that the price axis is very large and is discretized by the tick size (i.e. the minimum increment of price). The minimum 
and maximum price are $K$ tick away, with $K \gg 1$.
Time is measured in discrete steps, and in each time step, three types of events can happen: new limit orders are placed, market orders are placed or extant limit orders are canceled. 
Many of these events can happen simultaneously during a single time step.

\begin{itemize}
\item Limit orders are orders to buy or sell, that do not trigger an immediate trade. According to this, limit orders are placed on a support of size $K$ that
is in practice infinite, i.e. much larger than all other price scales in the system. Limit orders placed below the current midpoint price are 
considered as limit orders to buy, while those placed above the midpoint price are considered orders to sell. All limit orders have unit 
volume in the present version of the model (but this is in no way an essential ingredient). Limit orders arrive with a uniform rate $\lambda$
per unit time per unit price. In practice this means that in each time step, there is a probability $\lambda^n e^{-\lambda}/n!$ that $n$ new limit orders fall 
in each of the $K$ bins of the price axis. 
\item Market orders are orders that trigger an immediate transaction (with existing limit orders on the opposite best price level). They arrive
with a rate $\mu$ per unit time. The sign (direction) of market orders is generated using the algorithm proposed in \cite{LMF} with one active agent at any instant of time. The volume of 
market orders is chosen to be a random fraction $f$ of the opposite volume at the best where the distribution of $f$ is given by $P_\zeta(f)=\zeta (1-f)^{\zeta-1}$, 
where $\zeta$ is a parameter ($\zeta>0$), that determines the typical relative volume of market orders. When $\zeta \to 0$, market orders take all the prevailing 
volume on the opposite best.
\item In each time step, each limit order in the book has the probability $\nu_{\infty}$ of being cancelled. This Poisson process of 
cancellation defines the typical lifetime of a limit order (as $\nu^{-1}_{\infty}$).

\end{itemize}

The order of the processes in each time step is first the placement of limit orders,
then the possible execution of market orders, finally cancellation of some limit orders.
As shown in Fig. \ref{fig:diffusivity}, with the help of the parameters $\gamma$ and $\zeta$ we can tune the system to guarantee diffusive
prices for any value of the parameters $\mu, \lambda$ and $\nu_{\infty}$. This diffusive market allows us to test the predictions of our analytical model, 
about the locally linear profile around the current 
midpoint price. To do this, we add an extra ``agent'' to the model, who wishes to transact a metaorder that is typically larger than the 
available volume on the first level of the book and thus has to be split up and traded incrementally. It is the price change from the beginning
to the end of such a metaorder, that we would like to study, in order to compare the predictions of the model to empiricial results. The only 
remaining point is to define how the extra agent trades. In the present paper we allowed two trading strategies for the execution of
the metaorder: a) ``$\zeta$-execution'', where the extra agent trades 
exactly as the rest of the market, by sending a market order of volume $f \times q_\mathrm{best}$, where $f$ is chosen according to $P_\zeta(f)$ 
above and b) ``unit-execution'', where the market orders are all of unit volume, whenever he trades.

\begin{figure}
\begin{center}
\includegraphics[width=8cm, height=5.5cm]{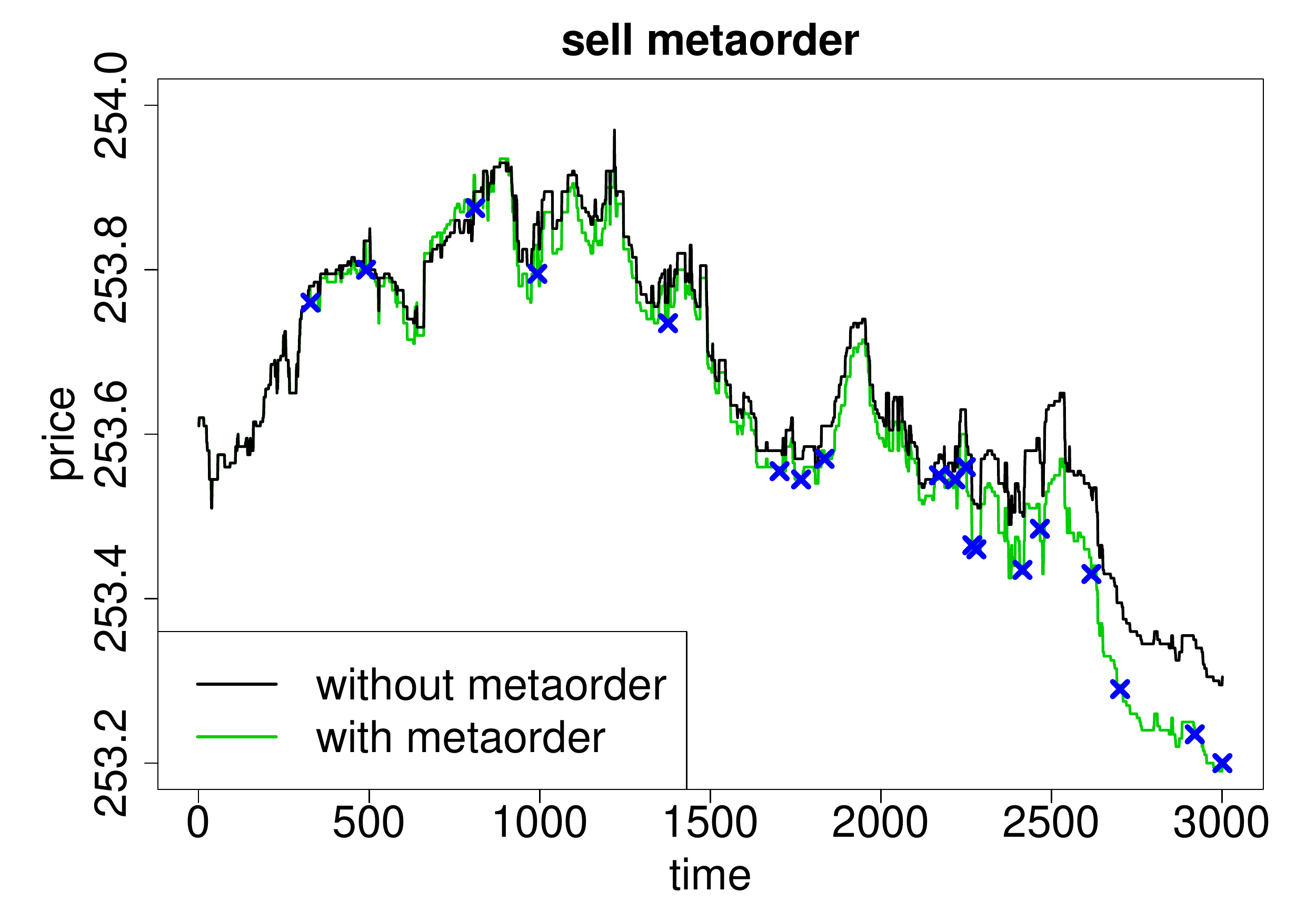}
\caption{\label{fig:app} The price process with and without the metaorder trading, with the same random seed used for the two simulations. The black curve shows tha price 
dynamics without the metaorder, while the green curve shows it with the metaorder, trades of the metaorder are denoted by blue crosses.
In the simulations we used the parameters, $\gamma=0.8$, $\zeta=0.65$, $\nu_\infty=10^{-4}$, $\lambda=0.5$. The participation rate of the metaorder is $\Phi=0.05$.}
\end{center}
\end{figure}

In Fig. \ref{fig:app} we show a segment of the price process for a simulation run, chosen only for illustration purposes. We plot the price as a function
of time (simulation steps). The two simulations shown are governed by the same random seed, however in one case an extra agent trading a metaorder {\it to sell}
is added to the market. The black curve shows the price dynamics without the metaorder, while the green curve shows it with the metaorder trades (the metaorder starts 
at time 0 and finishes at time 3000). Blue crosses indicate the moments when a trade by the extra agent was made. The agent follows the $\zeta$-execution strategy in this particular run.
For a while the two curves are identical, then after the first trade of the metaorder the two curves start to deviate from each other. The coarse-grained dynamics
of the two curves are similar, however the final price in case of the green curve is pushed down as a result of the extra sell metaorder.

\end{document}